\documentclass[11pt,a4paper]{article}
\usepackage{jheppub}
\usepackage[utf8]{inputenc} 
\usepackage{mathtools,xcolor,physics}
\hypersetup{citecolor=purple}

\newcommand{\chim}{{\langle \chi \rangle}}
\newcommand{\UV}{{\textrm{UV}}}
\newcommand{\IR}{{\textrm{IR}}}

\begin{document}

\title{Relevant Dilaton Stabilization}

\author[a]{Csaba Cs\'aki,}
\author[b]{Michael Geller,}
\author[b]{Zamir Heller-Algazi}
\author[a]{and Ameen Ismail}

\affiliation[a]{Department of Physics, LEPP, Cornell University, Ithaca, NY 14853, USA}
\affiliation[b]{School of Physics and Astronomy, Tel-Aviv University, Tel-Aviv 69978, Israel}

\emailAdd{csaki@cornell.edu}
\emailAdd{mic.geller@gmail.com}
\emailAdd{zamir.heller@gmail.com}
\emailAdd{ai279@cornell.edu}

\abstract{We propose a simple modification of the Goldberger-Wise mechanism for stabilizing the scale of spontaneously broken conformal theories. The source of explicit conformal symmetry breaking is a relevant operator with a small coefficient, as opposed to the usual mechanism of an almost marginal operator with an order-one coefficient. In the warped 5D picture this relevant stabilization corresponds to a small tadpole for the bulk scalar on the UV brane, which can be technically natural if it is the only source for the breaking of a symmetry (for example, a discrete $Z_2$). This modification of the stabilization mechanism has significant consequences for the nature of the conformal phase transition, since the radion/dilaton potential is no longer shallow. The bounce action is significantly reduced, leading to a weaker first-order phase transition instead of the supercooled and strongly first-order transition seen in Goldberger-Wise stabilization. This also leads to reduction of gravitational wave signals which, however, may still be observable at future detectors.
We present numerical and analytical studies of the phase transition and the resulting gravitational wave signal strength, assuming that the effective dilaton potential provides a good leading approximation. While the dilaton is not expected to be generically light in this setup, in order to keep  perturbative control over the effective theory one needs to mildly tune the dilaton quartic to be somewhat small.}

\arxivnumber{2301.10247}

\maketitle
\flushbottom

\section{Introduction}

One of the deepest mysteries of particle physics is the smallness of the observed Higgs mass~\cite{ATLAS:2012yve,CMS:2012qbp} and vacuum expectation value (VEV) compared to the scale where the Standard Model (SM) is expected to be completed into a fuller theory, which is na\"ively expected to lie around the Planck or GUT scale. Generically, the Higgs mass is expected to have a power-law dependence on the high scales at which any physics beyond the Standard Model manifests itself. Hence it is difficult to understand how the Higgs mass ended up being so much lighter, which is the well-known ``Higgs hierarchy'' or ``naturalness'' problem.

A commonly studied solution to this hierarchy problem is Higgs compositeness (see~\cite{Bellazzini:2014yua,Panico:2015jxa} for reviews and~\cite{Randall:1999ee,Randall:1999vf,Agashe:2004rs,Contino:2003ve,Rattazzi:2000hs,Arkani-Hamed:2000ijo,Csaki:1999mp,Tanaka:2000er,Csaki:2000zn,Goldberger:1999un,Goldberger:1999wh,Bellazzini:2013fga,Csaki:2022htl,Pomarol:2019aae,Csaki:2020zqz,Csaki:2017cep,Csaki:2015gfd,Blasi:2020ktl,Csaki:2008zd,Csaki:2007ns,Geller:2014kta,Agashe:2016rle,Sundrum:2009gv,Luty:2000ec,Luty:1999cz,Cline:1999ts,Agashe:2004cp,Gripaios:2009pe,Mrazek:2011iu,Giudice:2007fh,Contino:2006qr,Gherghetta:2000qt,Chacko:2014pqa,Chacko:2013dra,Chacko:2012sy,Batra:2008jy,Chacko:1999mi,Chacko:1999am,Barbieri:2015lqa} for studies in the field), where some new interaction becomes strong close to the weak scale, producing light composites including the Higgs boson itself, and thus providing a dynamical stabilization of the hierarchy.\footnote{A variation of this idea is Technicolor, in which the strong dynamics directly break the electroweak symmetry, similar to the dynamics of QCD. This idea, however, does not produce a light Higgs boson and is also expected to lead to large electroweak precision corrections, so it is strongly disfavored.} Composite Higgs models can also be studied via their holographic implementation~\cite{Agashe:2004rs,Contino:2003ve}, where the large hierarchy manifests itself as a warped extra dimension whose size is stabilized at large values and a corresponding exponentially small IR scale. In the simplest models proposed by Randall and Sundrum (RS)~\cite{Randall:1999ee,Randall:1999vf} the warped extra dimension is a slice of anti-de Sitter (AdS) space, spanning from the UV to the IR branes at its ends and corresponding to a near-conformal theory in the dual 4D implementation~\cite{Rattazzi:2000hs,Arkani-Hamed:2000ijo}. Fields localized close to the IR (including the Higgs boson) correspond to composites, while those in the UV are elementary (which usually includes some of the light SM fermions). The appearance of the IR brane can be interpreted as a spontaneous breaking of the conformal symmetry. The corresponding Goldstone boson is the radion excitation, associated to fluctuations of the IR brane~\cite{Csaki:1999mp,Tanaka:2000er,Csaki:2000zn,Goldberger:1999un}. In the 4D theory the radion is interpreted as the dilaton, the Goldstone boson of spontaneously broken scale invariance. The VEV of the dilaton/radion field sets the size of the warped extra dimension and consequently the IR scale. The solution to the hierarchy problem then comes down to the question of how to stabilize the dilaton/radion at large field values.

An elegant stabilization method was provided by Goldberger and Wise~\cite{Goldberger:1999uk}, who posited that a nearly marginal operator of dimension $4+\epsilon$ gets an expectation value. This leads to an exponentially large extra dimension in the 5D picture, through the exponential dependence of the dilaton VEV on $1/\epsilon$. It was first realized in~\cite{Creminelli:2001th} that this Goldberger-Wise stabilization mechanism has profound consequences for the early-universe behavior of these models~\cite{Randall:2006py,Agashe:2020lfz,Agashe:2019lhy,vonHarling:2017yew,Bruggisser:2022rdm,Baldes:2021aph,Bruggisser:2018mrt,Bruggisser:2018mus,Konstandin:2011dr,Baratella:2018pxi}. At high temperatures the conformal symmetry is restored, and the theory is essentially a hot conformal field theory (CFT). The holographic interpretation of this hot CFT is the modification of the AdS background to AdS-Schwarzschild --- a different solution to the same Einstein equations, corresponding to a black brane solution in 5D AdS space, with a black hole horizon at a finite proper distance from the UV brane and spanning the full 4D space~\cite{Creminelli:2001th}. The transition between the unbroken and broken CFT phases corresponds to the nucleation of bubbles of the IR brane. Goldberger-Wise stabilization yields a dilaton potential whose minimum is very shallow, resulting in a large contribution to the bounce action in the dilaton region, in turn making it difficult to complete the phase transition. This leads to the standard prediction that the RS phase transition is supercooled, strongly first-order, and often cannot complete until the temperature is well below the weak scale (the conditions for supercooled phase transitions were recently analyzed in detail in~\cite{Levi:2022bzt}). Another important consequence of Goldberger-Wise stabilization is on the spectrum of gravitational waves emitted during the phase transition~\cite{Randall:2006py}. Since the phase transition is strongly first order, one expects strong stochastic gravitational wave signals produced from bubble wall collisions. These could be detected at next-generation gravitational wave observatories, such as LISA~\cite{LISA:2017pwj,Baker:2019nia}, DECIGO~\cite{Seto:2001qf,Kawamura:2011zz,Yagi:2011wg,Isoyama:2018rjb}, and BBO~\cite{Crowder:2005nr,Corbin:2005ny,Harry:2006fi}. 

In this paper we point out that there is a simple variation of the Goldberger-Wise stabilization mechanism that would significantly alter the nature of the RS phase transition. Instead of having an almost marginal operator with small anomalous dimension obtain a VEV after a long running, one can have a relevant operator as the source of the spontaneous breaking of conformality. In this case the generation of a large hierarchy requires that the coefficient of this operator is very small in the UV, which can easily be made technically natural via a discrete symmetry. For example, in the Goldberger-Wise framework one can enforce a $Z_2$ symmetry for the bulk scalar, which is softly broken by a small tadpole on the UV brane. Deviating from regular Goldberger-Wise stabilization, in our ``relevant stabilization'' scenario the bulk mass of the bulk scalar is not small, and the hierarchy is generated by the smallness of the UV brane tadpole. This significantly alters the shape of the potential for the dilaton, such that it is no longer much lighter than the other Kaluza-Klein (KK) excitations. Since the potential is no longer very shallow, the bounce action is expected to be significantly reduced. This weakens the strength of the phase transition, allowing it to complete with no supercooling. The resulting gravitational wave  spectrum is peaked at a higher frequency and the overall signal strength is reduced.

Our goal in this work is to perform the first steps towards studying the nature of the RS phase transition with relevant stabilization. We will restrict ourselves to studying the dilaton effective action, assuming that it provides a reliable description of the theory. As expected, we will show that the bounce action is greatly reduced relative to the Goldberger-Wise case, making the phase transition weaker. We will also confirm that, within our approximation, the strength of the gravitational wave signal emitted during the phase transition is greatly reduced; it may still, however, be observable at future gravitational wave detectors. 

One other consequence of the relevant stabilization mechanism is that the calculability of the phase transition is also reduced, for two reasons. Firstly, since the bounce action in the calculable regime is smaller, it will no longer easily dominate over the (so far) noncalculable contribution of the hot phase. Secondly, since the dilaton is heavier, the gravitational and scalar KK modes might also become significant in the RS side of the phase transition. Both of these effects suggest that a more involved numerical study is necessary to firmly establish the results in this paper, where we only work in the limit of the dilaton effective action. We expect to address the phase transition in the full theory in subsequent work. 

The paper is organized as follows. We give a general overview of the relevant stabilization mechanism in Section~\ref{sec:general}. We then derive the effective dilaton potential in our model in Section~\ref{sec:potential}, showing that our dilaton is heavier than the Goldberger-Wise dilaton. We discuss preliminary aspects of the phase transition in Section~\ref{sec:phasetransition}, which sets the stage for detailed analytical and numerical calculations of the phase transition that we present in Section~\ref{sec:results}. We find that the gravitational wave signals from relevant stabilization are higher-frequency and weaker than those from Goldberger-Wise stabilization.

\section{The General Picture}\label{sec:general}

In this work, we are studying a composite Higgs scenario arising from the spontaneous breaking of a CFT, where the spontaneous breaking is triggered by a relevant operator
\begin{equation}
    \delta \mathcal{L}= g_d \mathcal{O},\quad \textrm{where }\bqty{\mathcal{O}}=d<4.
\end{equation}
If the coupling were ${\cal O}(1)$, then the presence of the operator would correspond to a large explicit breaking of the CFT, and no hierarchy can be generated, as any breaking scale generated would be not far below the UV scale. However, if the coupling $g_d$ is taken to be very small, then one still has an approximate CFT in the UV, and a large hierarchy can be created due to the running of this coupling. The form of the running of $g_d$ as a function of the renormalization scale $\mu$ is given by $g_d(\mu) = g_d\left(\mu_{\UV}\right)\left(\mu_{\UV}/\mu\right)^{4-d} $. Assuming that the coupling in the UV $g_d\left(\mu_{\UV}\right)$ is very small, the breaking of scale invariance is expected to be triggered when this running coupling becomes sizable at a scale $f$ far below the UV scale. This generates a UV/IR hierarchy that could be used for composite Higgs models in the usual way. 
  
One can then find the form of the effective dilaton potential by performing the usual spurion analysis, restoring scale invariance by treating the CFT breaking couplings as if they were operators with the correct scaling dimension. If $\mathcal{O}$ has dimension $d<4$, then the scaling dimension of its coupling $g_d$ is $4-d$. In order for the theory to be technically natural, we will also assume that $\mathcal{O}$ is odd under an additional $Z_2$ discrete symmetry. Using spurion analysis again, we assign the $g_d$ coupling to be odd under the same symmetry, which implies that only even powers of $g_d$ show up in the dilaton effective potential. Since $g_d^2$ has scaling dimension $8-2d$, there will exist a term in this potential wherein the dilaton field $\chi$ shows up with power $2d-4=2\nu$, in addition to the standard scale invariant $\chi^4$ term in the effective potential:
\begin{equation}
\label{eq: CFT_potential}
V_{\rm eff}\pqty{\chi}=\lambda \chi^4  -\lambda_{2\nu} \mu_\UV^{4-2\nu} \chi^{2\nu} +\ldots ,
\end{equation}
where we have introduced the dimensionless coupling $\lambda_{2\nu} \propto \gamma^2$, such that $\gamma \propto g_d\pqty{\mu_\UV} \mu_\UV^{d-4}$ is a dimensionless parameter characterizing the size of the explicit breaking of scale invariance in the UV. Note that this form of the potential is in agreement with~\cite{Rattazzi:2000hs}: while in the generic expression the lowest power appearing is $\chi^{2+\nu}$, this is absent for us due to our assumption of the additional $Z_2$ symmetry.\footnote{This can also be understood as assuming the CFT dynamics don't break the $Z_2$ symmetry, therefore $\ev{\mathcal O}$ must vanish with the explicit breaking source $g_d\pqty{\mu_\UV}$ and the potential has no linear piece in $g_d\pqty{\mu_\UV}$~\cite{Rattazzi:2000hs}.} We have also assumed that $\lambda>0$ and chosen the sign of the contribution of the explicit breaking term $\lambda_{2\nu}>0$, such that these two terms can balance each other to generate a stable minimum at $f=\chim = \mu_\UV \pqty{\nu\lambda_{2\nu}/2\lambda}^{1/(4-2\nu)}\sim\mu_\UV\gamma^{1/(2-\nu)} \ll \mu_{\UV} $. The smallness of the coupling in the UV therefore allows for a large hierarchy between the stabilized IR scale and the UV scale, as necessary to address the Higgs hierarchy problem.  
 
 One important question to address is whether a description in terms of a dilaton, corresponding to a mostly spontaneously broken scale invariance, is still valid, since we have introduced an explicit breaking to trigger the spontaneous breaking of scale invariance. When this description is invalid, no light dilaton with mass below the breaking scale $f$ should exist. Hence we find a self-consistency condition on the parameters of the potential in Eq.~\eqref{eq: CFT_potential}, which can be written as $\eval{V''(\chi)}_{\chi = f}/f^2 \lesssim \mathcal{O}\pqty{1}$. This will translate into the condition 
\begin{equation}\label{eq: CFT_dilaton-mass}
8 \pqty{2-\nu} \lambda \lesssim \mathcal{O}\pqty{1}.
\end{equation}

In the 5D picture, this scenario is realized by a bulk scalar field whose bulk mass term is negative (but not beyond the Breitenlohner-Freedman bound~\cite{Mezincescu:1984ev}), such that the profile of both solutions to the 5D Klein-Gordon equation grows from the UV to the IR brane. By the AdS/CFT dictionary the bulk mass $m^2$ is related to the dimension of the CFT operator by $d=2+\nu=2+\sqrt{4+m^2/k^2}$, where $k$ is the AdS curvature~\cite{Rattazzi:2000hs}. Unlike the Goldberger-Wise case, the bulk mass is not small in absolute size and the scalar has no VEV on the IR brane. In addition, a small UV brane tadpole is added for this scalar field, which is proportional to the explicit CFT breaking parameter $\gamma$ in the above CFT picture. As a result, a small VEV is generated on the UV brane, proportional to the size of the UV tadpole, in the absence of which the entire 5D VEV profile would vanish. As the VEV profile grows towards the IR, its contribution to the effective potential becomes comparable to that of the mistuning between the IR brane tension and the bulk CC, and balancing both terms stabilizes a large hierarchy without any small dimensions. Unlike the Goldberger-Wise potential, where the hierarchy is due to the small bulk mass corresponding to a small anomalous mass dimension $\epsilon = d-4$, in our scenario the hierarchy is directly generated by the technically natural small size of the tadpole.

There are several consequences of the absence of small dimensions in our scenario which make it distinct from the Goldberger-Wise case. First, the dilaton mass is of the same order as the IR scale, suppressed only by the CFT breaking coupling at the minimum $g_d\pqty{f}\sim \lambda$, where $\lambda$ is taken to be somewhat small for perturbativity.  This differs from the Goldberger-Wise limit, in which the dilaton mass is suppressed by the anomalous dimension $\epsilon^{1/2}$~\cite{Csaki:2000zn}. Next, the bounce action for the conformal phase transition is not enhanced by any small parameters, in contrast to Goldberger-Wise stabilization where the bounce action scales as $1/\epsilon^{3/4}$~\cite{Randall:2006py}. As a result, the phase transition is more weakly first-order and can easily complete without significant supercooling. This therefore alleviates the problem of eternal inflation, present in much of the parameter space of Goldberger-Wise stabilization~\cite{Creminelli:2001th} (see a modern expanded analysis in~\cite{Levi:2022bzt}). This also leads to a more rapid phase transition and thus weaker gravitational wave signatures with a higher peak frequency relative to the Goldberger-Wise case. We will see these effects clearly in Section~\ref{sec:results}.

Our proposed stabilization mechanism shares some similarities with that proposed in~\cite{Pomarol:2019aae}, where the operator $\mathcal{O}_g$ which breaks scale-invariance is marginal with a small imaginary anomalous dimension $2\sqrt{-\epsilon}$. Such operators can appear in CFTs where the IR fixed point merges with a UV fixed point, and the appearance of the imaginary anomalous dimension is interpreted as the loss of conformality. In the dual AdS picture, this can be realized by a bulk field with a mass below the Breitenlohner-Freedman bound, which therefore becomes tachyonic. The analysis of~\cite{Pomarol:2019aae} showed that in this scenario the dilaton mass is light but not parametrically smaller than the IR scale, similar to what is achieved in our model through a relevant operator. Interestingly, while in our model the heaviness of the dilaton renders the phase transition weakly first-order, the model of~\cite{Pomarol:2019aae} is believed to lead to strong supercooling.

\section{Dilaton Potential from the 5D Picture}\label{sec:potential}

We now calculate the effective dilaton potential of our stabilization mechanism in the 5D picture. The minimum of this potential will correspond to a hierarchically small dilaton VEV compared to the UV cutoff scale given by the AdS curvature $\mu_\UV=k$, thereby generating the UV/IR hierarchy needed to solve the naturalness problem.

We work in the RS background, with the metric
\begin{equation}\label{eq: RS metric}
    \dd{s}^2=e^{-2ky}\dd{x}^2-\dd{y}^2,
\end{equation}
where $y$ is the orbifolded fifth dimension. The UV and the IR branes are the two orbifold fixed points at $y=0$ and $y=y_c$, respectively. This metric is a solution to the Einstein equations when the bulk CC is $\Lambda=-24M_5^3k^2$ and the brane tensions are tuned to $\Lambda_\UV=-\Lambda_\IR=24M_5^3k$, where $M_5$ is the 5D Planck mass~\cite{Randall:1999ee}. 

We introduce a free scalar $\Phi$ in the bulk, whose action is given by
\begin{equation}\label{eq: stabilizing bulk matter}\begin{split}
	S_\Phi&=\int\dd[4]{x}\dd{y}\sqrt{g}\left[\frac{1}{2}g^{MN}\partial_M\Phi\partial_N\Phi-\frac{1}{2}m^2\Phi^2\right.\\
	&\hphantom{=\int\dd[4]{x}\dd{y}\sqrt{g} [ }\left.-\frac{\sqrt{g_{\rm ind}}}{\sqrt{g}}V_\UV\pqty{\Phi}\delta\pqty{y}-\frac{\sqrt{g_{\rm ind}}}{\sqrt{g}}V_\IR\pqty{\Phi}\delta\pqty{y-y_c}\right]
	\end{split}
\end{equation}
where $g_{\rm ind}$ is the determinant of the induced metric on the branes.
$\Phi$ respects a $Z_2$ symmetry which is softly broken by a small tadpole on the UV brane:
\begin{equation}
	V_\UV\pqty{\Phi}=\frac{1}{2}m_\UV\Phi^2+\gamma k^{5/2}\Phi,\qquad
	V_\IR\pqty{\Phi}=\frac{1}{2}m_\IR\Phi^2.
\end{equation}
$\gamma$ is dimensionless and can be taken to be very small since it is the only source of $Z_2$ breaking and thus technically natural. One way to generate a small $\gamma$ is via a Yukawa coupling $\Phi\psi_L\psi_R$, with one of the fermions odd under the $Z_2$ symmetry. A fermion condensate can be generated from the dynamics of some new confining gauge group similar to QCD. The scale of the condensate is controlled by dimensional transmutation and can be naturally smaller than the UV scale.

For the zero modes of $\Phi$, which are only $y$-dependent, the solutions to the bulk equations of motion (EOM) are $e^{\pqty{2\pm\nu}ky}$, where $\nu\equiv\sqrt{4+m^2/k^2}$. We assume that $0<\nu<2$ so that on the UV brane, the second solution dominates. The profile of $\Phi$ is localized towards the IR brane, and may be written as
\begin{equation}\label{eq: bulk scalar EOM}
	\Phi\pqty{y}=\Phi_0e^{\pqty{2-\nu}ky}\pqty{1+\Phi_1e^{2\nu ky}}.
\end{equation}
The boundary conditions (BCs) on the branes are
\begin{equation}\label{eq: BCs}
	2\Phi'\pqty{0}=m_\UV\Phi\pqty{0}+\gamma k^{5/2},\qquad
	-2\Phi'\pqty{y_c}=m_\IR\Phi\pqty{y_c}.
\end{equation}
This fixes the coefficients of the 5D VEV profile in Eq.~\eqref{eq: bulk scalar EOM} to be
\begin{equation}
	\Phi_1=-\frac{\tau_\IR e^{-2\nu ky_c}}{\tau_\IR+4\nu},\qquad
    \Phi_0=-\frac{\gamma k^{3/2}}{\tau_\UV+\Phi_1\pqty{\tau_\UV-4\nu}}\simeq-\frac{\gamma k^{3/2}}{\tau_\UV},
\end{equation}
where we defined the mass mistunings on the IR and UV branes as
\begin{equation}
\label{eq: mass_mistunings}
	\tau_\IR\equiv m_\IR/k+4-2\nu,\qquad
	\tau_\UV\equiv m_\UV/k-\pqty{4-2\nu}.
\end{equation}

The zero mode should be stable under small perturbations, otherwise the generated dilaton potential itself will also be unstable. We verify this by perturbing the solution $\Phi\pqty{y}+\phi\pqty{x^\mu,y}$ and plugging it back into $S_\Phi$. The effective 4D mass of the perturbation is obtained by solving the EOM for $\phi$ in the limit of small 4D momentum $p\ll ke^{-ky_c}$ and then integrating out the extra dimension, which is found to be $m_\phi ^2\propto\pqty{\tau_\UV+\tau_\IR e^{-2\nu ky_c}}k^2$. A positive UV mass mistuning $\tau_\UV$ therefore ensures the zero-mode perturbations are not tacyhonic and the solution is stable. 

As shown in App.~\ref{app. effective dilaton action}, the effective 4D potential of the dilaton is obtained by integrating out the bulk matter and substituting in the solutions to the EOM and BCs. Following this procedure with $S_\Phi$, integration by parts leaves us with only the boundary terms, since it is quadratic in $\Phi$:
\begin{equation}
\label{eq:dilaton_effective_action}
	V\pqty{\chi}=-\int\dd{y}\mathcal{L}_\Phi=-\Phi'\pqty{0}\Phi\pqty{0}+V_\UV+e^{-4ky_c}\pqty{\Phi'\pqty{y_c}\Phi\pqty{y_c}+V_\IR}.
\end{equation}
Once we impose the BCs in Eq.~\eqref{eq: BCs} we are left with the tadpole contribution
\begin{equation}
    V\pqty{\chi}=\frac{1}{2}\gamma k^{5/2}\Phi\pqty{0}=\frac{\tau_\IR\gamma^2}{2\tau_\UV\pqty{\tau_\IR+4\nu}}k^{4-2\nu}\chi^{2\nu}+\textrm{const.},
\end{equation}
where $\chi\equiv ke^{-ky_c}$ is the dilaton. This is the key piece of our potential --- the dilaton can have any power between $0$ and $4$ while the size of this coupling is proportional to $\gamma^2$, which can be hierarchically small. In the dual CFT this is the second term in Eq.~\eqref{eq: CFT_potential} which is quadratic in $g_d\pqty{\mu_\UV}$, as anticipated from the $Z_2$ symmetry of $\mathcal O$. No $\chi^{2+\nu}$ term is generated due to the vanishing VEV of $\Phi$ on the IR brane.

For comparison, in the Goldberger-Wise mechanism~\cite{Goldberger:1999uk} the scalar has a small bulk mass $\epsilon\equiv\sqrt{4+m^2/k^2}-2\ll1$ with nonzero VEVs on both branes, which induces couplings $\chi^{4+\epsilon},\chi^{4+2\epsilon}$ whose coefficients are $\mathcal{O}\pqty{1}$ in units of $k$. It is dual to an almost marginal operator of dimension $4+\epsilon$ where conformal invariance is broken both explicitly and spontaneously~\cite{Rattazzi:2000hs}. However, as pointed out in~\cite{Rattazzi:2000hs}, it is not necessary for the bulk scalar to obtain a VEV on the IR brane for the Goldberger-Wise mechanism to work, in which case only the $\chi^{4+2\epsilon}$ term is generated. Our mechanism is similar to Golberger-Wise stabilization of this second variety with $\epsilon<0$.

To complete our construction of the dilaton potential, we mistune the IR and UV tensions. The former will induce a $\chi^4$ term in the potential, and the latter will give a constant term. The entire dilaton potential is therefore
\begin{equation}\label{eq: entire dilaton potential}
    V\pqty{\chi}=\frac{24M_5^3}{k^3}\bqty{\lambda\chi^4-\lambda_{2\nu}k^{4-2\nu}\chi^{2\nu}+V_1}.
\end{equation}
The overall scaling is added for later convenience. In our 5D realization, the coefficient of the first term is given by 
\begin{equation}\label{eq: explicit relevant coupling coefficient}
    \lambda_{2\nu}=-\frac{k^3}{48M_5^3}\frac{\tau_\IR}{\tau_\UV\pqty{\tau_\IR+4\nu}}\gamma^2.
\end{equation}
The existence of a non-trivial minimum requires $\lambda_{2\nu}>0$, so the IR mass mistuning must lie in the range $0>\tau_\IR>-4\nu$.

This potential admits a minimum at
\begin{equation}
    \chim=k\pqty{\frac{\lambda_{2\nu}\nu
    }{2\lambda}}^{1/\pqty{4-2\nu}}\sim k \gamma^{1/\pqty{2-\nu}}.
\end{equation}
A small value for $\ev{\chi}/k$ is generated by the technically natural $\gamma$, even when the power is not large. This is the novel feature of our stabilization mechanism: a large hierarchy of scales is generated without any small operator dimensions, but rather by a small explicit breaking of the CFT. This is unlike the Goldberger-Wise mechanism, where the large hierarchy is due to the smallness of $\epsilon$ while the scalar VEVs on the branes are of the same order. Note that as $\nu$ approaches $2$ our stabilizing operator becomes almost marginal and $1/\pqty{2-\nu}$ grows large, so $\gamma$ no longer needs to be small to generate a large hierarchy. In this limit our mechanism coincides with the Goldberger-Wise one with anomalous dimension $\epsilon=\nu-2$.

We set $V\pqty{\chim}\approx 0$, a tuning corresponding to the standard CC problem which is not addressed in this work. The dilaton potential is then
\begin{equation}\label{eq:dilatonpotentialfinal}
    V\pqty{\chi}=\frac{3N^2\lambda}{2\pi^2}\chim^4\bqty{\pqty{\chi/\chim}^4-1-\frac{\pqty{\chi/\chim}^{2\nu}-1}{\nu/2}}.
\end{equation}
Here we used the holographic relation $N^2=16\pi^2\pqty{M_5/k}^3$. With the kinetic term, the dilaton action is
\begin{equation}
    S_\chi=\int\dd[4]{x}\bqty{\frac{3N^2}{4\pi^2}\pqty{\partial\chi}^2-V\pqty{\chi}}.
\end{equation}

\begin{figure}
    \centering
    \includegraphics[width=0.8\textwidth]{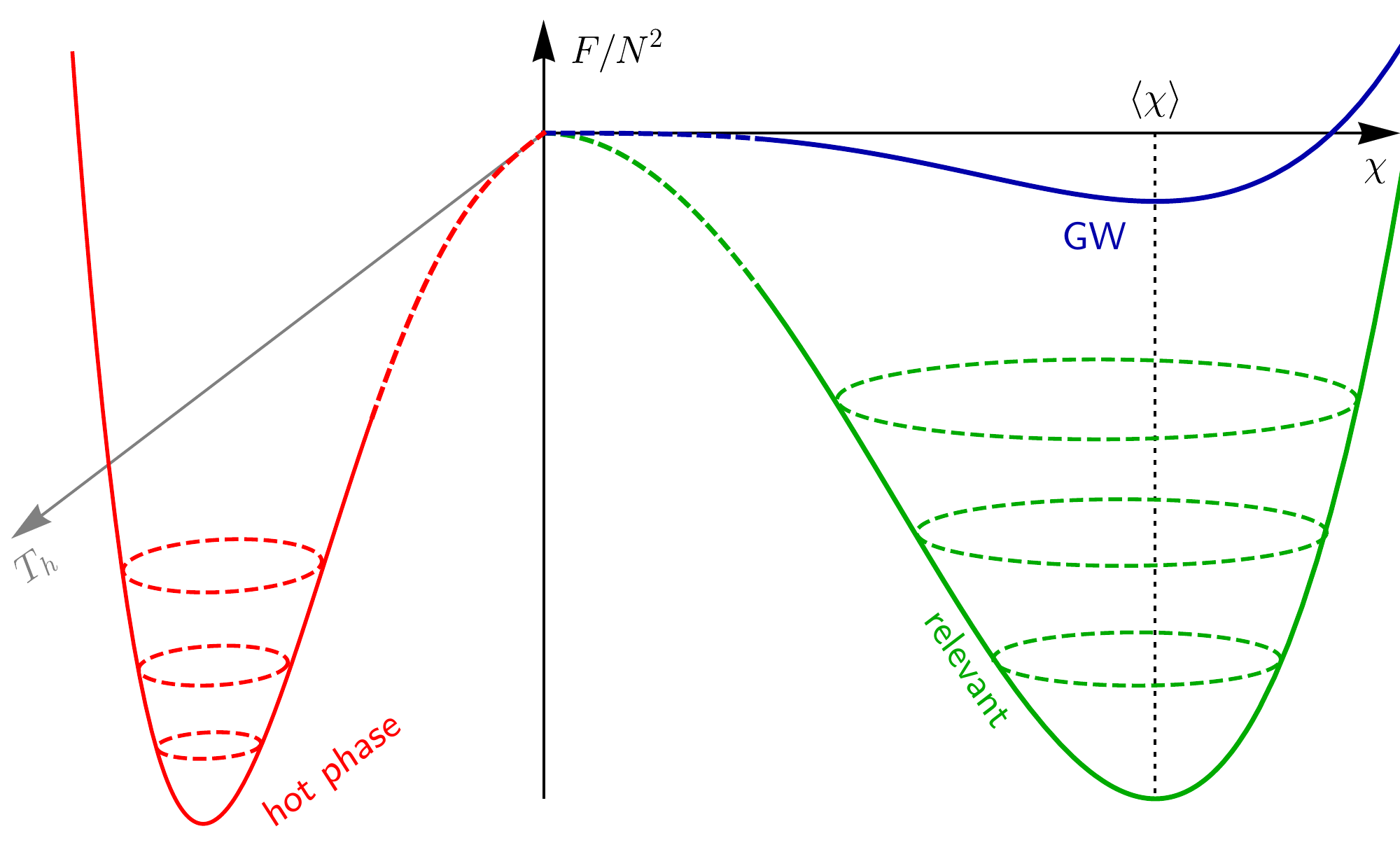}
    \caption{A sketch of the free energy of the CFT. The free energy of the cold, confined phase is given by the dilaton potential in RS, with the Goldberger-Wise mechanism in \textit{blue} and the relevant stabilization mechanism we propose in \textit{green}. On the left side the free energy of the hot, deconfined phase is given by the AdS-S metric, as a function of the horizon temperature $T_h$ \textit{(red)}~\cite{Creminelli:2001th}.}
    \label{fig: RS phase diagram}
\end{figure}

The mass of the dilaton is given by
\begin{equation}\label{eq: dilaton mass}
    m_\chi^2=\frac{2\pi^2}{3N^2}\eval{\pdv[2]{V}{\chi}}_{\chi=\chim}=8\lambda\pqty{2-\nu}\chim^2,
\end{equation}
which for relevant operators is of the same order as the IR scale. In contrast, as the operator becomes almost marginal ($\nu \rightarrow 2$) the dilaton's mass is suppressed by the small anomalous dimension, as is the case in the Goldberger-Wise mechanism~\cite{Csaki:2000zn}. The dilaton potential in the relevant stabilization mechanism is steeper than for the Goldberger-Wise stabilization, as illustrated in Fig.~\ref{fig: RS phase diagram}. Because the dilaton mass is no longer suppressed, the graviton~\cite{Randall:1999vf,Tanaka:2000er} and scalar~\cite{Goldberger:1999wh,Csaki:2000zn} KK modes might become significant. To ensure the lowest-lying KK modes are not excited, we will require that $m_\chi\lesssim M_{\rm KK}$, as anticipated in Eq.~\eqref{eq: CFT_dilaton-mass} in the CFT picture.

Our analysis of the stabilization mechanism was done entirely within the effective dilaton theory, which would break down if the backreaction on the metric is too large. We estimate this backreaction is small when the potential on the IR brane, $V_\IR\pqty{\Phi}+24\lambda M_5^3k$, is smaller than the IR brane tension $\Lambda_\IR$. This condition also ensures that the CFT breaking is spontaneous. While the CFT-breaking coupling blows up for a small enough $\chi$, i.e. below $\chi_*$ where $V(\chi_*)/\chi_*^4=4\pi $ (the potential is dominated by the CFT breaking term), at $\chi = \chim$ both couplings balance one another and are of the same order. Therefore, at $\chi = \chim$, for small enough $\lambda$ the CFT-breaking term is still perturbative, and correspondingly the bulk scalar field does not significantly backreact on the AdS background, which is broken spontaneously by the appearance of the IR brane.

In the following, we will explicitly study our model for the benchmark point $\nu=1.2,m_\IR=-4k$ with the two UV mass mistunings $\tau_\UV=3,10$. The first mode of the stabilizing scalar is the lightest KK mode with $M_{\rm KK}=2.1\chim$, independent of the value of $\tau_\UV$ (the mass of the first graviton KK mode is larger, $\approx3.8\chim$). The backreaction on the metric is small when $\lambda\lesssim0.13$ $\pqty{\lambda\lesssim0.6}$ for $\tau_\UV=3$ $\pqty{\tau_\UV=10}$, which also guarantees the dilaton is lighter than the KK modes. Thus the validity of the dilaton EFT requires a mild tuning of the quartic.
 
\section{Phase Transition}\label{sec:phasetransition}

As the universe cools down, the CFT undergoes a phase transition from the hot deconfined phase to the cold confined phase. In the dual 5D picture, the hot phase is described by a black brane solution to the Einstein equations; in the limit where the UV brane is taken to the AdS boundary this solution is just the AdS-Schwarzschild metric.\footnote{With the UV brane, the solution is no longer stationary. It corresponds to a radiation dominated universe which expands with time, which can be thought of as a moving UV brane~\cite{Gubser:1999vj,Kraus:1999it}.} The cold phase is dual to the usual RS picture with an IR brane. The phase transition proceeds via nucleation of IR brane bubbles within the black brane background~\cite{Creminelli:2001th}.

The critical temperature of the phase transition $T_c$ is determined by matching the free energies of the confined and deconfined phases. The former is given by the dilaton effective potential in Eq.~\eqref{eq:dilatonpotentialfinal}, $F_{\rm conf}(\chi) \approx V(\chi)$ (for $T\lesssim M_{\rm KK}(\chi)$); the latter is $F_{\rm deconf}(T) = -\pi^2 N^2 T^4/8+ V_0$. The constant term $V_0$ can be found by identifying a common limit to the two phases~\cite{Creminelli:2001th,Agashe:2020lfz}, leading to $V_0 = \frac{3N^2 \lambda (2-\nu)}{2\pi^2\nu}\chim^4 $. Solving for $F_{\rm conf}(\chim) = F_{\rm deconf}(T_c)$, the critical temperature is thus
\begin{equation}
    T_c = \frac{\chim}{\pi} \left [ 12\lambda \frac{2-\nu}{\nu} \right]^{1/4} .
\end{equation}
We remark that the free energy of the deconfined phase can be written as $F_{\rm deconf}(T) = \pi^2 N^2 (T_c^4 - T^4) / 8$.

The phase transition proceeds when the bubble nucleation rate $\Gamma \sim T^4 e^{-S_b}$, where $S_b$ is the Euclidean bounce action, is larger than the Hubble parameter $H^4$. For $T < T_c$ we can approximate
\begin{equation}
    H^2 \approx \frac{F_{\rm deconf}(T=0)}{3 M_{\rm Pl}^2} = \frac{\pi^2 N^2 T_c^4}{24 M_{\rm Pl}^2} .
\end{equation}
This leads to an upper bound on the bounce action for the phase transition to complete:
\begin{equation}\label{eq:bouncecrit}
    S_b \lesssim 4 \log \frac{M_{\rm Pl}}{T_c} .
\end{equation}
Thus for a TeV-scale dilaton VEV and critical temperature, the phase transition does not complete until $S_b \lesssim 140$~\cite{Agashe:2019lhy,vonHarling:2017yew}. In part of the parameter space for the standard Goldberger-Wise mechanism, this condition is never satisfied and the universe is stuck in eternal inflation with a positive CC in the deconfined phase. In our scenario, this doesn't occur for sufficiently large $\lambda$.

First-order phase transitions can lead to stochastic gravitational wave signatures resulting from bubble wall collisions. The strength of the gravitational wave signal is controlled by the inverse duration of the phase transition $\beta_{\rm GW}$, which is approximately given by~\cite{Caprini:2015zlo,Caprini:2019egz}
\begin{equation}\label{eq:GWstrength}
    \frac{\beta_{\rm GW}}{H} = T \eval{\dv{S_b}{T}}_{T=T_n}
\end{equation}
where $T_n$ is the nucleation temperature at which the phase transition occurs, and $H$ is the Hubble parameter at $T = T_n$. The gravitational wave signal strength is inversely proportional to $(\beta_{\rm GW}/H)^2$~\cite{Caprini:2015zlo,Caprini:2019egz}.

The dynamics of the phase transition are controlled by the bounce solution which interpolates between the vacua of the two phases~\cite{Coleman:1977py,Linde:1981zj}. However, the dilaton effective theory of the confined phase breaks down when the temperature is larger than the mass of the lightest KK mode, $T > M_{\rm KK} \sim \chim$ (see Fig.~\ref{fig: RS phase diagram} and~\cite{Creminelli:2001th}), and part of the bounce occurs in the noncalculable regime of the deconfined phase. A computation of the dynamics of the phase transition performed entirely within the 4D dilaton EFT will therefore always have some theoretical uncertainty.
This uncertainty is greater for our scenario than in the conventional Goldberger-Wise case for the two reasons already mentioned in the introduction: our dilaton is heavier, so the effect of the KK modes is more important; and since the calculable portion of the bounce is much smaller, it does not necessarily dominate over the noncalculable part.

In principle, it is possible to obtain the exact bounce solution by solving the full 5D EOM --- which are Euclidean-time Einstein equations with two coordinates, the radial direction of the bounce and the direction of the extra dimension --- and in fact this has been done numerically in a similar scenario in~\cite{Aharony:2005bm}. This has yet to be accomplished for the black brane-RS bounce, and is beyond the scope of this paper. In what follows we will work entirely within the dilaton effective theory despite this inherent uncertainty. Nevertheless, we will see that we can still make useful predictions with regard to the bounce action (Fig.~\ref{fig:bounceaction}), nucleation temperature (Fig.~\ref{fig:GW}), and gravitational wave signals (Fig.~\ref{fig:exclusion}).

Since we cannot obtain the exact bounce solution of the full 5D theory, we instead use a proxy trajectory in the effective 4D theory. Following~\cite{Creminelli:2001th}, we glue the free energies of the confined and deconfined phases at $\chi=0$, taking the dilaton potential to be $F_{\rm conf}(\chi)$ for $\chi>0$ and equal to $F_{\rm deconf}(T)$ for $\chi<0$. Note that there is a discontinuity in this potential at $\chi = 0$, which grows with $T$ and vanishes as $T \rightarrow 0$. We then consider bounce configurations $\chi(r)$ interpolating between the vacua $\chi = 0$ and $\chi = \chim$, with the boundary conditions $\chi'(0) = 0$ and $\chi(\infty) = 0$.

\section{Results}\label{sec:results}

\subsection{Thin-wall Analysis}

It is instructive to first analyze the bounce in the thin-wall approximation~\cite{Coleman:1977py,Linde:1981zj}. This is a good approximation when the temperature $T$ is close to $T_c$, and provides useful intuition for the bounce more generally.

The $O(3)$-symmetric bounce action\footnote{The $O(4)$-symmetric bounce action is always larger in the thin-wall limit, so the bounce is dominated by $O(3)$-symmetric bubbles.} is given by~\cite{Linde:1981zj}
\begin{equation}
    S_b = \frac{16\pi}{3} \frac{S_1^3}{\Delta V^2 T} ,
\end{equation}
where $\Delta V = \pi^2 N^2 (T_c^4 - T^4) / 8$ is the potential difference between the ends of the bounce, and $S_1$ is the bubble wall surface tension. The surface tension is determined by the potential as
\begin{equation}
    S_1 = \int_0^\chim \dd{\chi} \sqrt{2 V(\chi)} .
\end{equation}
Using the potential in Eq.~\eqref{eq:dilatonpotentialfinal}, we then find the bounce action at leading order in the expansion parameter $\delta = 1 - T/T_c$:
\begin{equation}\label{eq:thinwallbounce}\begin{split}
    S_b &\approx \frac{N^2}{3^{1/4} \lambda^{3/4} \delta^2} F(\nu)^3, \\
    F(\nu) &= \left( \frac{\nu}{2-\nu} \right)^{3/4} \int_0^1 \dd{x} \sqrt{ \frac{1 - x^{2\nu}}{\nu/2} - \left(1- x^4 \right) } .
\end{split}\end{equation}
The thin-wall approximation is valid when $\delta \ll 1$.

We remark that the bounce action scales as $N^2/\lambda^{3/4}$ and is independent of $\chim$, which remains true outside of the thin-wall limit. Furthermore, the bounce is not enhanced further by any small parameters, unlike the case of Goldberger-Wise stabilization where the bounce is enhanced by the small explicit breaking of scale invariance~\cite{Randall:2006py}. Because of this, it is possible for the phase transition to complete promptly, i.e. without supercooling where $T \ll T_c$.

Using Eq.~\eqref{eq:GWstrength}, the inverse duration of the phase transition in the thin-wall approximation is
\begin{equation}\label{eq:thinwallGW}
    \frac{\beta_{\rm GW}}{H} = \frac{2}{\delta} S_b .
\end{equation}
Note this is to be evaluated at the nucleation temperature, at which $S_b \approx 140$, as explained below Eq.~\eqref{eq:bouncecrit}.

\subsection{Numerics}

Next we compare the thin-wall results above to numerical calculations of the phase transition computed using the \texttt{FindBounce} package~\cite{Guada:2020xnz}. The usual boundary conditions need to be modified for the purpose of the computation due to the aforementioned potential discontinuity. The effect of the discontinuity can be absorbed into a modification of the boundary condition $\chi(\infty) = 0$ to $\chi'(r_*)^2 / 2 = \delta V$, where $r_*$ is the point at which $\chi(r_*) = 0$ and $\delta V$ is the size of the discontinuity. The bounce solution for $r > r_*$ is simply $\chi(r) = 0$.

\begin{figure}
    \centering
    \includegraphics[width=0.8\textwidth]{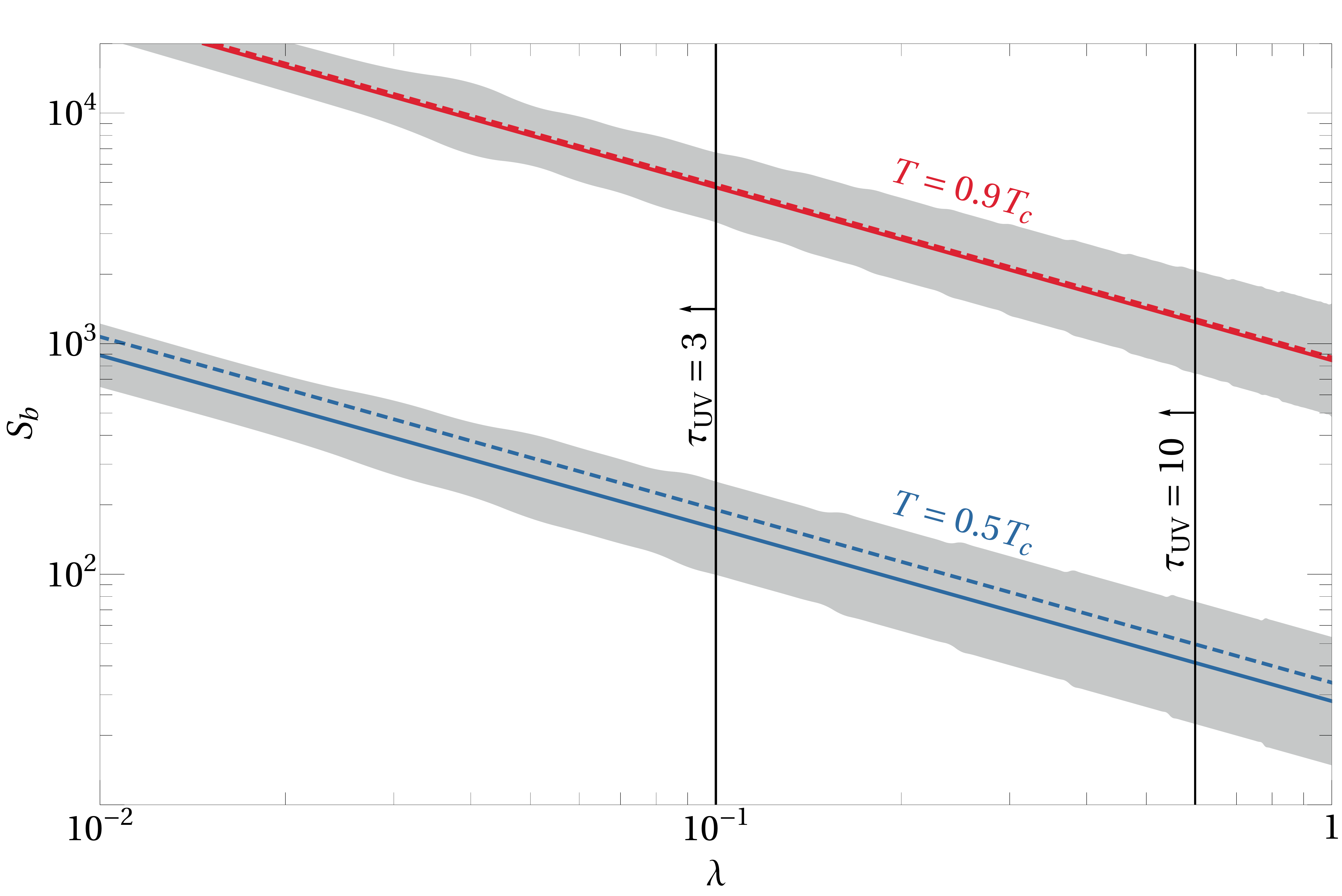}
    \caption{Comparison of the thin-wall approximation and numerical results for the $O(3)$-symmetric bounce action $S_b$. The solid lines are numerical calculations and the dashed lines correspond to the thin-wall approximation. We fix $\nu = 1.2$ and $N = 5$ and choose $T = 0.5T_c$ \textit{(blue lines)} and $T = 0.9 T_c$ \textit{(red lines)}, as the quartic $\lambda$ is varied from $0.01$ to $1$. The inherent error arising from the breakdown of the dilaton effective theory is depicted by the gray bands. We also include the value of $\lambda$ at which the backreaction of the bulk scalar on the metric becomes large \textit{(black lines)} for $\tau_{\rm UV} = 3$ and $\tau_{\rm UV} = 10$.}
    \label{fig:bounceaction}
\end{figure}

As a check of our numerical methods, Fig.~\ref{fig:bounceaction} depicts the bounce action as a function of the quartic $\lambda$ for $T = 0.5 T_c$ ($\delta = 0.5$) and $T = 0.9 T_c$ ($\delta=0.1$), fixing $\nu=1.2$ and $N=5$. We show our numerical computations alongside the thin-wall result in Eq.~\eqref{eq:thinwallbounce}. The $1/\lambda^{3/4}$ scaling is manifest. As expected, the thin-wall approximation and the numerical result are in excellent agreement for $\delta = 0.1$, and are of the same order of magnitude when $\delta = 0.5$.

Recall that there is a theoretical uncertainty in calculating the bounce action within the 4D dilaton effective theory, as part of the bounce occurs in the noncalculable regime. To estimate the error we scale the potential by a constant $V \rightarrow (1+\varepsilon) V$ in the noncalculable regime $0 < \chi < T(\chi/M_{\rm KK})$, then compute the rate of change of the bounce action under this scaling, $\dv*{S_b}{\varepsilon}$. We then take the relative error in the bounce action to be $\abs{S_b^{-1} \dv*{S_b}{\varepsilon}}$, which characterizes the sensitivity of the bounce action to the noncalculable regime. This error is depicted as shaded bands in Fig.~\ref{fig:bounceaction}. It should be understood as a crude estimate rather than a rigorous computation of the theoretical uncertainty, which would require a more involved analysis in the dual 5D picture. Lastly, we show the value of $\lambda$ at which the scalar field has a significant backreaction on the metric as black lines in Fig.~\ref{fig:bounceaction}. For $\tau_{\rm UV} = 3$ ($\tau_{\rm UV} = 10$) we need $\lambda \lesssim 0.1$ ($\lambda \lesssim 0.6$) to ensure a small backreaction.

\begin{figure}
    \centering
    \includegraphics[width=0.8\textwidth]{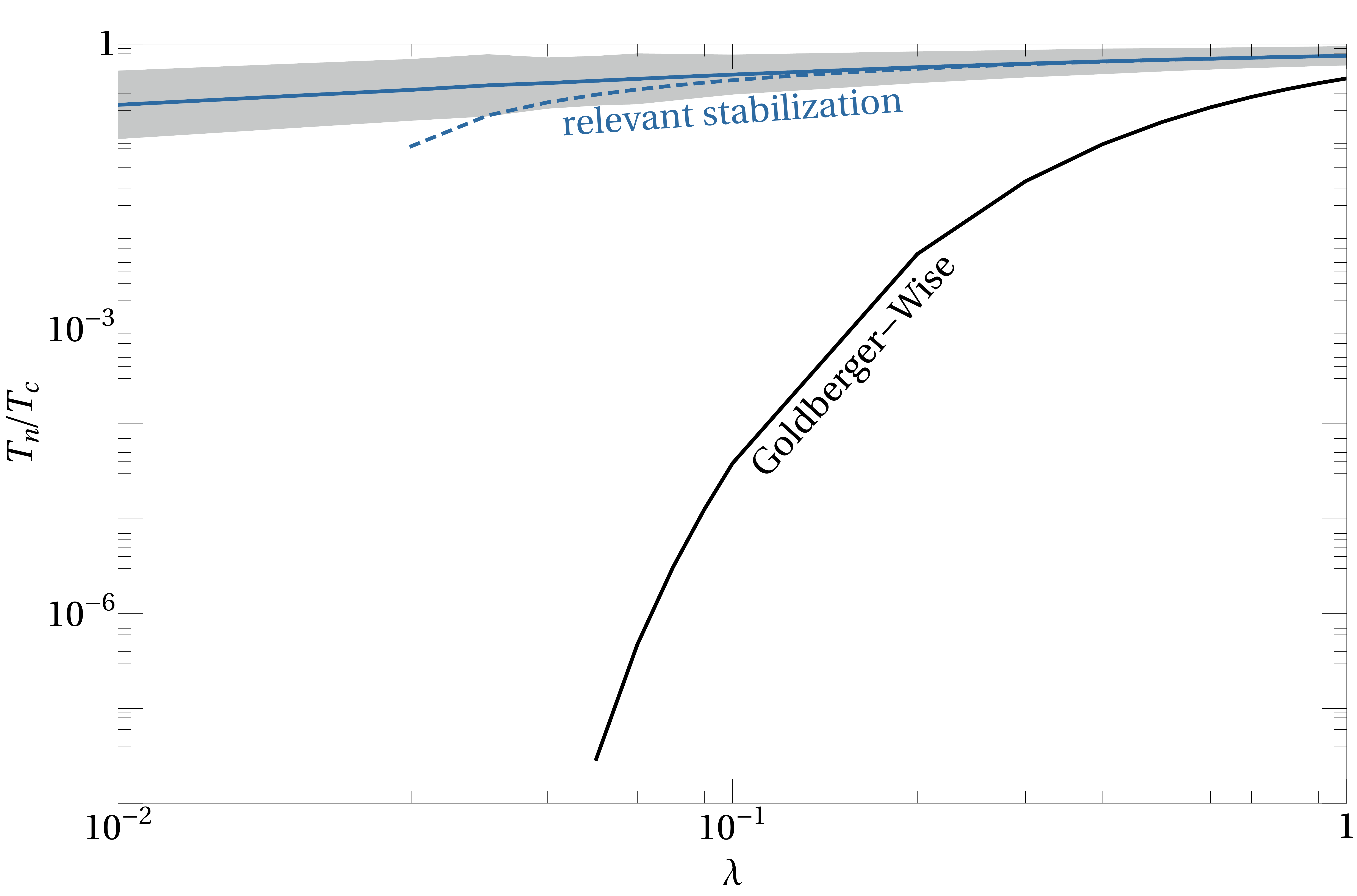}\\
    \includegraphics[width=0.8\textwidth]{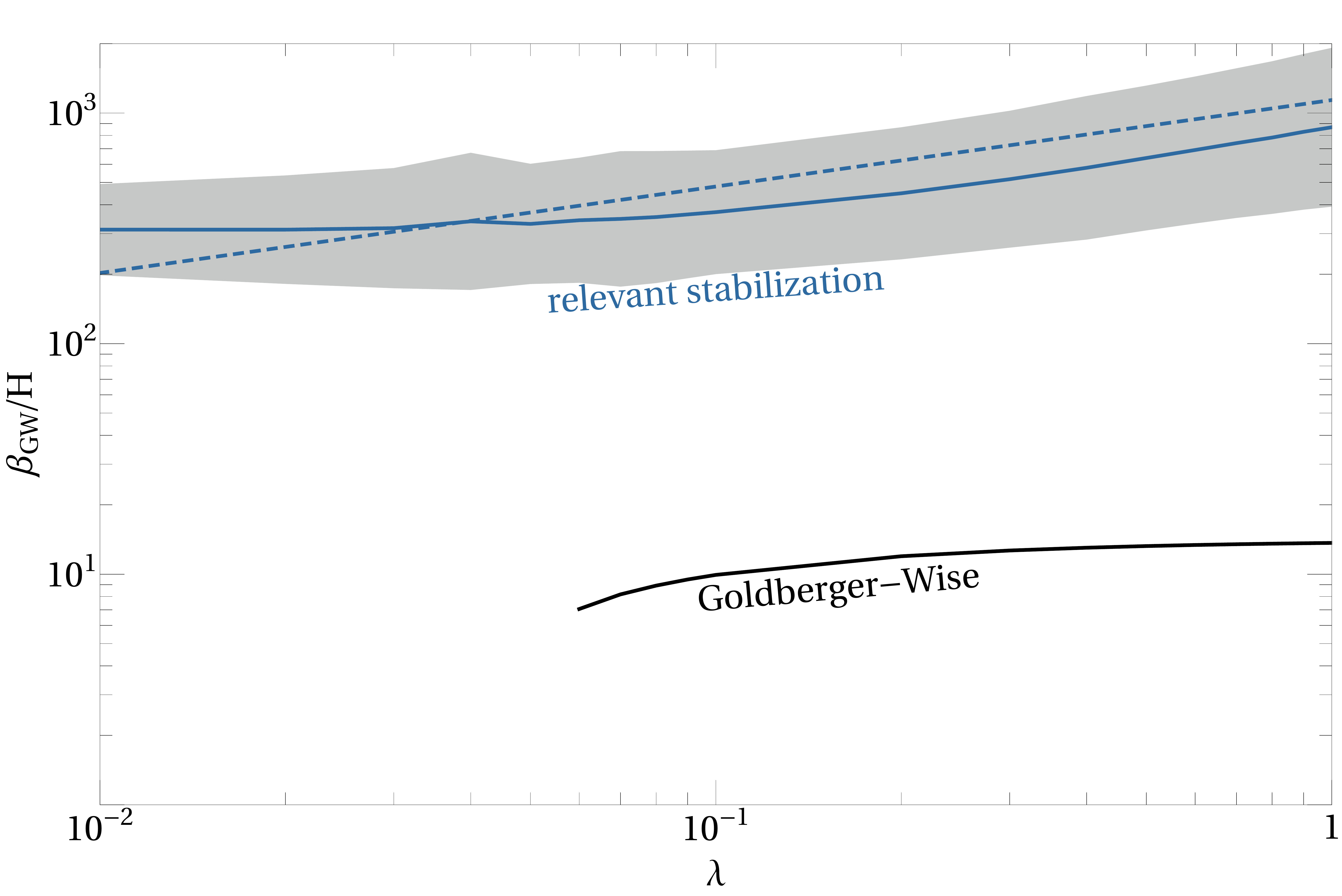}
    \caption{The ratio of the nucleation temperature to the critical temperature $T_n/T_c$ \textit{(top)} and the ratio of the inverse duration of the phase transition to the Hubble parameter $\beta_{\rm GW}/H$ at the time of transition  \textit{(bottom)}. We fix $\nu = 1.2$ and $N = 5$. The solid blue lines are computed numerically, the dashed blue lines are computed in the thin-wall approximation, and the gray bands estimate the theoretical error due to the breakdown of the dilaton effective theory. For comparison we include the corresponding values for a Goldberger-Wise stablized dilaton with $\epsilon=-1/20$ and $N = 5$ \textit{(black line)}.}
    \label{fig:GW}
\end{figure}

\begin{figure}
    \centering
    \includegraphics[width=0.8\textwidth]{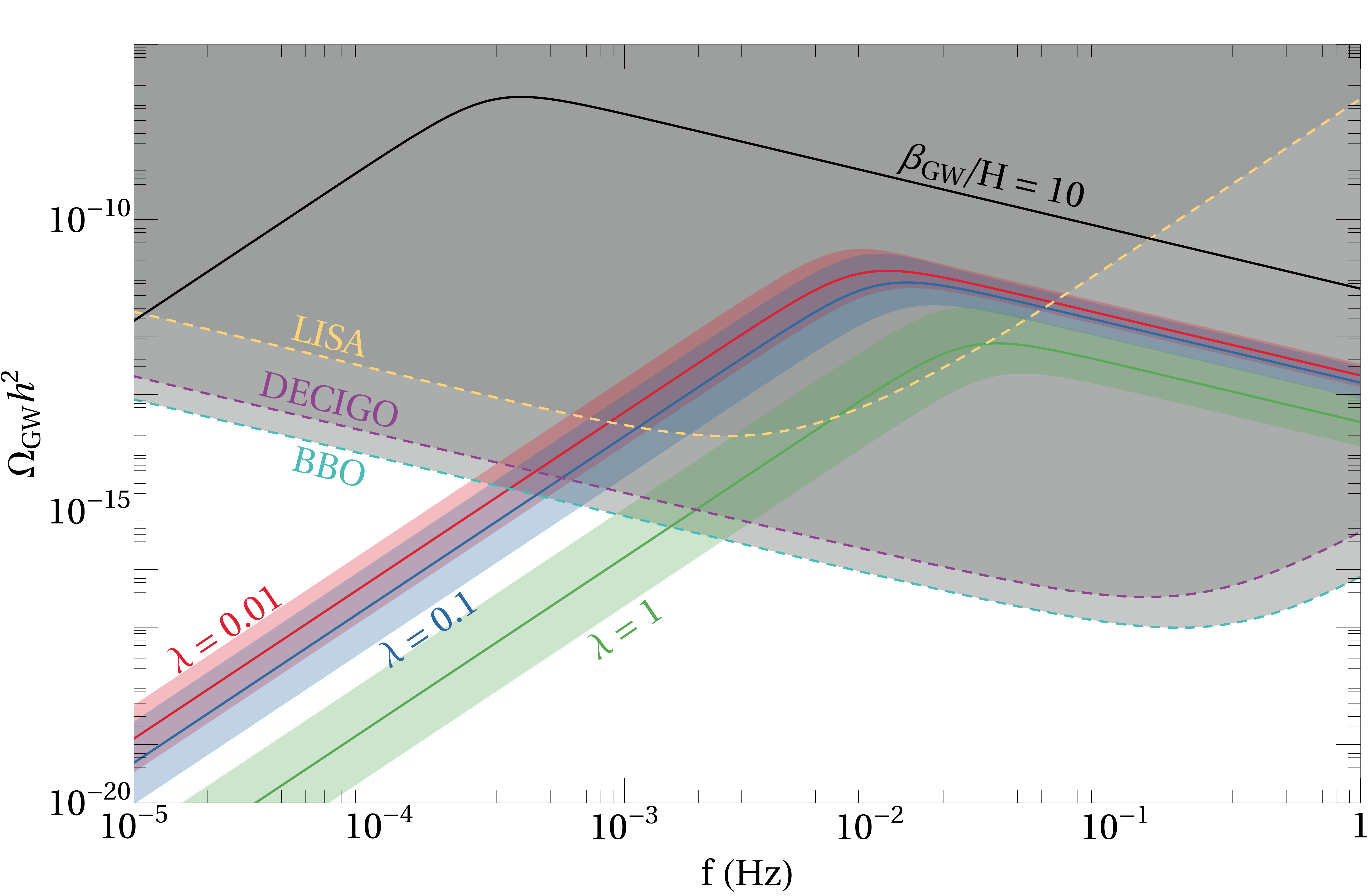}
    \caption{The gravitational wave abundance spectrum $\Omega_{\rm GW} h^2(f)$ for $\lambda = 0.01$ (\textit{red}), $\lambda = 0.1$ (\textit{blue}), and $\lambda = 1$ (\textit{green}), fixing $\nu = 1.2$ and $N = 5$. The colored bands indicate the theoretical error due to the dilaton EFT breaking down. For comparison we include a spectrum for $\beta_{\rm GW}/H = 10$ (\textit{black}), a typical value for Goldberger-Wise stabilization. We show projected experimental sensitivities for LISA~\cite{LISA:2017pwj,Baker:2019nia} (\textit{orange, dashed}), DECIGO~\cite{Seto:2001qf,Kawamura:2011zz,Yagi:2011wg,Isoyama:2018rjb} (\textit{purple, dashed}), and BBO~\cite{Crowder:2005nr,Corbin:2005ny,Harry:2006fi} (\textit{turquoise, dashed}).}
    \label{fig:exclusion}
\end{figure}

Our main results are contained in Figs.~\ref{fig:GW} and~\ref{fig:exclusion}. Fig.~\ref{fig:GW} shows the nucleation temperature $T_n$ and the inverse duration of the phase transition $\beta_{\rm GW}$ in relevant stabilization. We again fix $\nu = 1.2$ and $N = 5$ and present both numerical computations and the thin-wall approximation. We also depict the sensitivity to the noncalculable regime with shaded bands again. The thin-wall result for $T_n$ is obtained by setting $S_b = 140$ in Eq.~\eqref{eq:thinwallbounce} and solving for $T/T_c$; we then use Eq.~\eqref{eq:thinwallGW} to compute $\beta_{\rm GW}$. Although we present results for the quartic $\lambda$ ranging from $10^{-2}$ to $1$, recall that for large values the dilaton EFT is no longer valid. The point the EFT breaks down depends on the 5D model parameters as depicted in Fig.~\ref{fig:bounceaction}.

For comparison we consider a Goldberger-Wise stabilized dilaton with $\epsilon = -1/20$ (corresponding to $\nu = 2 - 1/20$), computing $T_n$ and $\beta_{\rm GW}$ in the thick-wall approximation~\cite{Linde:1981zj} following~\cite{vonHarling:2017yew}, which is a good approximation in the supercooled limit. In contrast to the Goldberger-Wise case, our mechanism requires no substantial supercooling for the phase transition to complete. Consequently, the inverse duration of the phase transition $\beta_{\rm GW}/H$ is larger for our model, of order $10^2$ or $10^3$, whereas for the Goldberger-Wise stabilized dilaton $\beta_{\rm GW}/H \sim 10$ is typical. This will lead to weaker gravitational wave signals in our model.

Fig.~\ref{fig:exclusion} contains gravitational wave spectra computed using our numerical results for the phase transition duration. We depict the gravitational wave abundance $\Omega_{\rm GW} h^2$ as a function of frequency $f$ for $\lambda = 10^{-2},10^{-1},1$, as well as a spectrum for $\beta_{\rm GW}/H = 10$, which was what we found for Goldberger-Wise stabilization in Fig.~\ref{fig:GW}. We also show projected sensitivities for three proposed gravitational wave detectors in Fig.~\ref{fig:exclusion} --- LISA~\cite{LISA:2017pwj,Baker:2019nia}, DECIGO~\cite{Seto:2001qf,Kawamura:2011zz,Yagi:2011wg,Isoyama:2018rjb}, and BBO~\cite{Crowder:2005nr,Corbin:2005ny,Harry:2006fi} --- as computed in~\cite{Schmitz:2020syl} assuming a signal-to-noise ratio of 1. We assume the signal arises entirely from bubble collisions, that is, we ignore contributions from sound waves and turbulence. We model the bubble collisions using the envelope approximation, reviewed in~\cite{Caprini:2015zlo,Caprini:2019egz}, under the following assumptions: the bubble wall velocity is $1$, the effective number of degrees of freedom during the phase transition is $g_* = 100$, and the temperature immediately after the phase transition is $1$~TeV. In App.~\ref{sec:GWdetails} we justify our approximations and provide explicit formulae for $\Omega_{\rm GW} h^2$. We note that for $\lambda=1$ the dilaton EFT cannot be trusted, so the results for this benchmark point should be interpreted with caution.

The gravitational wave signals in our model are weaker by several orders of magnitude than in Goldberger-Wise stabilization and are shifted towards higher frequencies. As explained above, this is due to the lack of supercooling and relatively weak first-order phase transition. Nevertheless, one could still probe all of our benchmark points at DECIGO and BBO, and all but possibly the $\lambda = 1$ point at LISA.

\section{Conclusions}

In this work we have described a new way to stabilize the scale of spontaneously broken conformal symmetry. Instead of a nearly marginal operator acquiring a VEV like the Goldberger-Wise mechanism, in our mechanism a relevant operator with a small, technically natural coefficient gets a VEV. The small coefficient of the relevant operator generates a large UV/IR hierarchy. We calculated the effective dilaton potential in the dual 5D picture, and found that our dilaton typically has a mass of the same order as the IR scale, in contrast to the Goldberger-Wise dilaton whose mass is suppressed by the small anomalous dimension. One consequence of the enhanced dilaton mass is that a mild tuning of the dilaton quartic will be required to ensure the validity of our dilaton EFT.

Working within the dilaton effective theory, we studied the dynamics of the conformal phase transition. Our analytical approximations in the thin-wall limit as well as our numerical studies generally confirm our intuition about the phase transition: the bounce action is reduced relative to the Goldberger-Wise case because the dilaton potential is deeper. Thus, the phase transition is far more weakly first-order and proceeds without substantial supercooling. The major phenomenological effect resulting from this is that the stochastic gravitational wave signals from bubble collisions are reduced. However, they may still be observable at the next generation of gravitational wave detectors.

We emphasize that our use of the 4D dilaton EFT impedes the precision of our calculations. We cannot trust the dilaton potential near the origin, where the effective theory breaks down, and also part of the bounce occurs in the deconfined phase which is noncalculable. Although we have attempted to characterize the theoretical uncertainty in our computations, a complete treatment of the phase transition would require working in the full 5D picture and solving the (Euclidean-time) Einstein equations for the bounce configuration. We intend to study the phase transition from a 5D perspective more rigorously in future work.

\acknowledgments{
CC and AI are supported in part by the NSF grant PHY-2014071. CC is also supported in part by the US-Israeli BSF grant 2016153.  AI is also supported in part by NSERC, funding reference number 557763. MG is supported in part by Israel Science Foundation under Grant No. 1302/19. MG is also supported in part by the US-Israeli BSF grant 2018236 and NSF-BSF grant 2021779.
}

\appendix

\section{Derivation of the Effective Dilaton Action}\label{app. effective dilaton action}

In this appendix we derive the general effective dilaton action in the 5D picture. Similar results to ours were obtained in~\cite{Bellazzini:2013fga,Csaki:2022htl,Pomarol:2019aae}. We consider the RS action~\cite{Randall:1999ee} with additional matter,
\begin{equation}
    S=-\int\dd[4]{x}\dd{y}\bqty{\sqrt{g}\pqty{2M_5^3R+\Lambda}+\sqrt{g_{\rm ind}}\Lambda_\UV\delta\pqty{y}+\sqrt{g_{\rm ind}}\Lambda_\IR\delta\pqty{y-y_c}}+S_\textrm{m},
\end{equation}
where $R$ is the 5D Ricci scalar, the bulk CC and boundary tensions are taken to their RS values (see Sec.~\ref{sec:potential}) and $S_\textrm{m}$ is the action of additional matter in the bulk. We add scalar perturbations to the RS metric in Eq.~\eqref{eq: RS metric} using the following ansatz~\cite{Csaki:2000zn},
\begin{equation}\label{eq: RS+pert metric ansatz}
	\dd{s}^2=e^{-2\pqty{A+F}}\eta_{\mu\nu}\dd{x}^\mu\dd{x}^\nu-\pqty{1+2F}^2\dd{y}^2.
\end{equation}
$A(y)$ is the warp factor and $F(x^\mu,y)$ are the scalar perturbations, which we parameterize as $F(x^\mu,y)=f(y)r(x^\mu)$ and identify $r(x^\mu)$ as the radion. When $S_\textrm{m}$ and $F$ are taken to zero, the background solution of the Einstein equations is $A=ky$.

Working to leading order in the backreaction $\delta A\pqty{y}$, or equivalently in $\pqty{k/M_5}^3$, we note that the $T^{\textrm{m}}_{MN}$ calculated from $S_\textrm{m}$ can be taken at zeroth order. This follows from the Einstein equations, $G_{MN}=\frac{1}{4M_5^3}T_{MN}$, where evidently $G_{MN}$ is already first order in $\pqty{k/M_5}^3$, leaving $T^{\textrm{m}}_{MN}= T^{\textrm{m},(0)}_{MN}$. $T^{\textrm{m},(0)}_{MN}$ is calculated from $S_\textrm{m}=S^{(0)}_\textrm{m}(y)$. From the 4D Lorentz invariance of $S^{(0)}_\textrm{m}(y)$, it follows that $T^{\textrm{m},(0)}_{\mu 5}=0$, and the leading order of the $\pqty{\mu5}$ Einstein equation reads
\begin{equation}
	3\partial_\mu r\pqty{f'-2A'f}=0.
\end{equation}
Its solution gives the well-known radion profile $f\pqty{y}=e^{2A}$~\cite{Charmousis:1999rg}. Plugging in this profile to the rest of the Einstein equations, the $\pqty{55}$ component is then
\begin{equation}\label{eq: warp factor perturbation EOM}
	12k\delta A'+3e^{4A}\Box r=\frac{1}{4M_5^3}T_{55}^{\textrm{m},(0)}.
\end{equation}
Note that in the limit of no backreaction and no matter fields, Eq.~\eqref{eq: warp factor perturbation EOM} is the EOM of a massless radion field, as expected in this limit where the radion is not stabilized. The $\pqty{\mu\nu}$ components of the EOM include singular pieces in $\delta A''$, which impose the Israel junction conditions
\begin{equation}\label{eq: backreaction BCs}
	\eval{2\eta_{\mu\nu}e^{-2A}\delta A'}_{y=0,y_c}=\eval{\pm \frac{1}{12M_5^3}T_{\mu\nu}^{\textrm{m},(0)}}_{y=0,y_c}.
\end{equation}

We can now compute the effective dilaton action. Its minimum is obtained by solving $\fdv*{S}{r}=0$, which corresponds to solving the Einstein equations, imposing the BCs in Eq.~\eqref{eq: backreaction BCs}, as well as solving the EOM of the bulk matter fields in $S_\textrm{m}$. Therefore, in the vicinity of the minimum, the effective dilaton action is given by
\begin{equation}
	S_\textrm{eff}\pqty{r}=\int \fdv{S}{r}\dd{r}.
\end{equation}

By varying the action $S$ with respect to $r$ we obtain the Einstein equations,
\begin{equation}
	\fdv{S}{r}=\pqty{\fdv{S_\textrm{EH}}{g^{MN}}+\fdv{S_\Lambda}{g^{MN}}+\fdv{S_\textrm{m}}{g^{MN}}}\fdv{g^{MN}}{r}
	=\int\dd[4]{x}\dd{y}\sqrt{g}\pqty{-2M_5^3G_{MN}+\frac{1}{2}T_{MN}}\fdv{g^{MN}}{r}.
\end{equation}
We now plug in the metric ansatz in Eq.~\eqref{eq: RS+pert metric ansatz} to leading order and impose the bulk EOM, but we do not yet impose the Israel junction conditions in Eq.~\eqref{eq: backreaction BCs}. This allows us to calculate the effective action for the dilaton field away from its minimum, where the IR brane jump condition is satisfied. The UV brane jump condition will be equivalent to the requirement that the total 4D CC is zero, which we will assume to be satisfied. This leaves us with
\begin{equation}
    \fdv{S}{r}=\int\dd[4]{x}\sqrt{g}\bqty{\mp12M_5^3\eta_{\mu\nu}\delta A' e^{-2A}+\frac{1}{2}T_{\mu\nu}^{\textrm{m},(0)}}_{y=0,y_c} \fdv{g^{\mu\nu}}{r}.
\end{equation}
Indeed, we see that a minimum of the action is obtained once the BCs in Eq.~\eqref{eq: backreaction BCs} are satisfied. We substitute $\delta A'$ from Eq.~\eqref{eq: warp factor perturbation EOM} and obtain
\begin{equation}\label{eq: midstep derivation}
    \fdv{S}{r}=\int\dd[4]{x}\bqty{-\frac{24M_5^3}{k}\Box r\pqty{e^{2ky_c}-1}+\frac{2}{k}\eval{T^{\textrm{m},(0)}_{55}e^{-2A}}_{0}^{y_c}+\frac{1}{2}\eval{\sqrt{g}T_{\mu\nu}^{\textrm{m},(0)}\fdv{g^{\mu\nu}}{r}}_{y=0,y_c}}.
\end{equation}
The first term in this equation is the variation of the kinetic term of the radion~\cite{Csaki:2000zn,Goldberger:1999un}. The remaining terms, as we now show, are precisely $\fdv*{S_\textrm{m}^{(0)}}{r}$ to leading order: varying \textit{only} the matter fields $S_\textrm{m}^{(0)}$ with respect to $r$ gives
\begin{equation}
	\fdv{S_\textrm{m}^{(0)}}{r}=\int\dd[4]{x}\dd{y}\frac{1}{2}\sqrt{g}T^{\textrm{m},(0)}_{MN}\fdv{g^{MN}}{r}
    =\eval{\fdv{S_\textrm{m}^{(0)}}{r}}_{\rm{bulk}}+\frac{1}{2}\int\dd[4]{x}\eval{\sqrt{g}T_{\mu\nu}^{\textrm{m},(0)}\fdv{g^{\mu\nu}}{r}}_{y=0,y_c}.
\end{equation}
We separated out the contribution of the singular terms on the branes from the contribution of the smooth part of the bulk. The latter can be shown to be equal to the second term in Eq.~\eqref{eq: midstep derivation},
\begin{equation}
	\eval{\fdv{S_\textrm{m}^{(0)}}{r}}_{\textrm{bulk}}
    =\int\dd[4]{x}\dd{y}e^{-4A}f\pqty{4T^{\textrm{m},(0)}_{55}+2 e^{2A}\eta^{\mu\nu}T^{\textrm{m},(0)}_{\mu\nu}}
    =\int\dd[4]{x}\frac{2}{k}\eval{T^{\textrm{m},(0)}_{55}e^{-2A}}_0^{y_c},
\end{equation}
where we used the energy-momentum conservation relation
\begin{equation}
	0=\nabla^M T^{\textrm{m},(0)}_{M5}=-\partial_5T^{\textrm{m},(0)}_{55}+4A'T^{\textrm{m},(0)}_{55}+A'\eta^{\mu\nu}T^{{\textrm{m},(0)}}_{\mu\nu}.
\end{equation}

In total, we find that
\begin{equation}
	\fdv{S}{r}=-\int\dd[4]{x}\frac{24M_5^3}{k}\Box r\pqty{e^{2ky_c}-1}+\fdv{S_\textrm{m}^{(0)}}{r},
\end{equation}
and upon integrating we see that the effective dilaton action is given by
\begin{equation}\label{eq: effective dilaton potential}
   S_\textrm{eff}\pqty{\chi}=\frac{12M_5^3}{k^3}\int\dd[4]{x}\partial_\mu\chi\partial^\mu\chi+S_{\rm{m,eff}}^{(0)}.
\end{equation}
Here we reparametrized the radion as the dilaton,
\begin{equation}
    \chi\pqty{x}\equiv k \exp(-ky_c-r\pqty{x}e^{2 k y_c}).
\end{equation}
We have found that the effective dilaton potential is given by integrating the bulk matter action over solutions to the EOM (including appropriate BCs). The contribution of the backreaction is already encoded in Eq.~\eqref{eq: effective dilaton potential}. We use this calculation of the effective action in Eq.~\eqref{eq:dilaton_effective_action} in the main text. 

\section{Gravitational Wave Spectrum}\label{sec:GWdetails}

Here we provide an explicit expression for the gravitational wave abundance and carefully consider the assumptions which go into it. The reader is referred to~\cite{Caprini:2015zlo,Caprini:2019egz} for a pedagogical review of gravitational waves from first-order phase transitions.

The gravitational wave spectrum arises from three main processes: collisions of bubble walls, sound waves in the plasma, and turbulence in the plasma. We assumed that the contribution from bubble wall collisions dominates. Whether this is a good assumption depends on the ratio of vacuum energy density released in the phase transition to the energy density of the radiation bath. For us this is given by
\begin{equation}
    \alpha = \frac{15 N^2}{4 g_*} \left( \frac{T_c^4}{T_n^4} - 1 \right),
\end{equation}
where $g_*$ is the number of effective relativistic degrees of freedom during the phase transition. When $\alpha$ is large relative to a characteristic value $\alpha_\infty$, the sound wave and turbulence contributions can be safely neglected. Explicitly $\alpha_\infty$ is given by a sum over the masses of the particles that acquire a mass during the phase transition:
\begin{equation}
    \alpha_\infty = \frac{30}{24\pi^2 g_* T_n^2} \sum c_i m_i^2,
\end{equation}
where the $i$-th particle has mass $m_i$ after the transition and $c_i$ ($2 c_i$) degrees of freedom for bosons (fermions).

During the phase transition, the techni-quarks of the CFT sector confine into mesons. Assuming that the meson masses are all of the order of the dilaton VEV $\chim$, one can then calculate the ratio $\alpha/\alpha_\infty$ for the benchmark points in Fig.~\ref{fig:exclusion}. We find that $\alpha/\alpha_\infty$ is always larger than $1$ as long as there are less than about 200 mesonic degrees of freedom. In this case it is justified to neglect the sound wave and turbulence contributions to the gravitational wave spectrum.

Furthermore, in the $\alpha \gg \alpha_\infty$ limit, all of the energy released in the phase transition contributes to accelerating the bubble walls (as opposed to the bulk motion of the fluid) and the bubble wall velocity approaches the speed of light. Using the envelope approximation, the gravitational wave abundance from bubble wall collisions is then given by
\begin{equation}\label{eq:GWspectrum}
    \Omega_{\rm GW} h^2 = 1.3 \times 10^{-6} \left( \frac{H}{\beta_{\rm GW}} \frac{\alpha}{1+\alpha} \right)^2 \left( \frac{100}{g_*} \right)^{1/3} \frac{3.8 (f/f_p)^{2.8}}{1 + 2.8 (f/f_p)^{3.8}},
\end{equation}
where
\begin{equation}
    f_p = 3.8 \times 10^{-5} {\rm ~Hz~} \frac{\beta_{\rm GW}}{H} \frac{T}{1{\rm ~TeV}} \left( \frac{g_*}{100} \right)^{1/6}
\end{equation}
is the frequency the abundance is peaked at. The signal curves in Fig.~\ref{fig:exclusion} were computed using Eq.~\eqref{eq:GWspectrum}.

\bibliographystyle{JHEP}
\bibliography{references}

\end{document}